\documentclass[12pt,preprint]{aastex}

\shorttitle{}
\shortauthors{}
\slugcomment{Draft \today}

\begin{document}

\newcommand{\php}[0]{\phantom{--}}
\newcommand{\kms}[0]{km~s$^{-1}$}

\title{OBSERVATIONS OF MASS LOSS FROM THE TRANSITING EXOPLANET HD~209458b
\footnote{Based on observations made with the NASA/ESA {\it Hubble Space 
Telescope}, obtained from the Data Archive at the Space Telescope 
Science Institute. STScI is operated by the Association of Universities for 
Research in Astronomy, Inc., under NASA contract NAS 5-26555. These 
observations are associated with the {\it HST} GTO program \#11534.}}

\author{Jeffrey L. Linsky}
\affil{JILA, University of Colorado and NIST, 440 UCB Boulder, CO 80309-0440}
\email{jlinsky@jilau1.colorado.edu}

\author{Hao Yang}
\affil{JILA, University of Colorado and NIST, 440 UCB Boulder, CO 80309-0440}
\email{haoyang@jilau1.colorado.edu}

\author{Kevin France}
\affil{CASA, University of Colorado, 593 UCB Boulder, CO 80309-0593}
\email{Kevin.France@colorado.edu}

\author{Cynthia S. Froning}
\affil{CASA, University of Colorado, 593 UCB Boulder, CO 80309-0593}
\email{cfroning@casa.colorado.edu}

\author{James C. Green}
\affil{CASA, University of Colorado, 593 UCB Boulder, CO 80309-0593}
\email{James.Green@colorado.edu}

\author{John T. Stocke}
\affil{CASA, University of Colorado, 593 UCB Boulder, CO 80309-0593}
\email{stocke@casa.colorado.edu}

\author{Steven N. Osterman}
\affil{CASA, University of Colorado, 593 UCB Boulder, CO 80309-0593}
\email{steven.osterman@colorado.edu}

\begin{abstract}

\noindent Using the new Cosmic Origins Spectrograph (COS) 
on the {\it Hubble Space Telescope (HST)}, 
we obtained moderate-resolution, high signal/noise
ultraviolet spectra of HD 209458 and its 
exoplanet HD 209458b during transit, both orbital quadratures, and secondary 
eclipse. We compare transit spectra with spectra 
obtained at non-transit phases to identify spectral 
features due to the exoplanet's expanding atmosphere. We find that the 
mean flux decreased by $7.8\pm 1.3$\% for the C~II 1334.5323~\AA\ and 
1335.6854~\AA\ lines and by $8.2\pm 1.4$\% for
the Si~III 1206.500~\AA\ line during transit compared 
to non-transit times
in the velocity interval --50 to +50 km~s$^{-1}$. 
Comparison of the C~II and Si~III line depths and transit/non-transit line 
ratios shows deeper absorption features near --10 and +15 km~s$^{-1}$ and 
less certain features near --40 and +30--70 km~s$^{-1}$, 
but future observations are needed to verify this first detection of velocity
structure in the expanding atmosphere of an exoplanet.
Our results for the C~II
lines and the non-detection of Si~IV 1394.76~\AA\ absorption 
are in agreement with \citet{Vidal-Madjar2004},
but we find absorption during transit in the Si~III line contrary to the
earlier result. The $8\pm 1$\% obscuration
of the star during transit is far larger than the 1.5\% obscuration 
by the exoplanet's disk. Absorption during transit at velocities 
between --50 and +50~km~s$^{-1}$ in the C~II and Si~III lines requires
high-velocity ion absorbers. Assuming hydrodynamic 
model values for the gas temperature and outflow velocity at the limb 
of the outflow as seen in the C~II lines, we find mass-loss rates
in the range (8--40)$\times 10^{10}$ g~s$^{-1}$. These rates assume that
the carbon abundance is solar, which is not the case for the giant planets
in the solar system.
Our mass-loss rate estimate is consistent with theoretical hydrodynamic 
models that include metals in the outflowing gas.

\end{abstract}


\section{INTRODUCTION}

\citet{Mazeh2000} monitored the radial velocities of the G0~V star HD~209458 
to derive the orbital parameters and minimum mass
of its transiting planet HD 209458b. Very shortly thereafter, 
accurate measurements of the mass and 
radius of HD~209458b were extracted from radial velocity and photometric 
transit observations by \citet{Henry2000} and \citet{Charbonneau2000}.
Table 1 lists the presently accepted properties of HD~209458 
and HD~209458b cited by \citet{Knutson2007}. 
HD~209458b is likely the best-studied Jupiter-like 
exoplanet located very close to its host star. As a result of its proximity, 
HD~209458b receives very strong incident radiation (see estimate in Table~1) 
and stellar wind flux from its host star.

HD 209458b is the first transiting planet for which atmospheric absorption
was observed in the resonance lines of Na~I \citep{Charbonneau2002} and 
subsequently in lines of H~I, C~II, O~I, and others. 
Using the G140M grating of Space Telescope Imaging Spectrograph (STIS) on 
the Hubble Space Telescope {\it HST} with its resolution of 
$\sim 30$ km~s$^{-1}$, 
\citet{Vidal-Madjar2003} found that the Lyman-$\alpha$ line flux was 
reduced by $15\pm 4$\% during transit at velocities 
between --130 and +100 km~s$^{-1}$. Based on
this first detection of H~I absorption at or above the planet's Roche lobe,
they concluded that hydrogen is escaping from the planet.  
\citet{Vidal-Madjar2004} then used the
low-resolution G140L grating of STIS to 
detect an absorption depth at midtransit of
$5\pm 2$\% for the unresolved Lyman-$\alpha$ line, 
$13\pm 4.5$\% for the O~I 1304~\AA\ multiplet, 
and $7.5\pm 3.5$\% for the C~II 1335~\AA\ doublet. Their detection of
absorption in the O~I and C~II lines at or above the Roche lobe confirmed
significant mass loss from the planet by hydrodynamic outflow.
They argued that reduction in the Lyman-$\alpha$ line flux at 
$\approx 100$ km~s$^{-1}$ from line center indicates high velocities
of neutral hydrogen atoms. Additionally, more absorption on the blue side of 
the Lyman-$\alpha$ emission line than the red side 
indicates outflow towards the observer and away from the planet.   
A subsequent observation of HD~209458 using the {\it HST} Advanced
Camera for Surveys (ACS) showed a reduction in the unresolved 
Lyman-$\alpha$ line flux of ($8.0\pm 5.7$)\% during transits, 
consistent with the STIS G140L and G140M results, thereby 
confirming mass loss from the planet's large exosphere \citep{Ehrenreich2008}.

There have been a number of reanalyses of the data and futher discussions of 
the STIS transit G140M observations by \citet{Ben-Jaffel2007},  
\citet{Ben-Jaffel2008}, \citet{Vidal-Madjar2008} and others. There is a 
consensus that HD~209458b has an extended exosphere and is losing mass, but
important details are not yet understood. Such questions as: whether
the exosphere size is larger or smaller than the projected Roche lobe, what 
are the outflow speed and mass-loss rate, and whether or not the mass loss 
from the planet produces a cometlike tail are not yet answered 
with the existing STIS and ACS observations. 
While the STIS G140M Lyman-$\alpha$ spectrum obtained by
\citet{Vidal-Madjar2003} provides some velocity information, higher spectral 
resolution and signal/noise (S/N) are needed to 
obtain the velocity structure and optical depth of the outflowing gas in 
Lyman-$\alpha$ and other spectral lines. 
 
The large abundance of hydrogen and large flux in the Lyman-$\alpha$ line make
this line a prime candidate for studying exoplanet outflows, but there are 
uncertainties (many of them unique)  
in measuring the exoplanet's mass-loss rate only from this line. 
These include the modest S/N of the existing 
STIS G140M grating data, possible variability 
of the stellar Lyman-$\alpha$ profile at the time of transit that 
cannot be directly measured, 
broad interstellar absorption in the core of the line that prevents
measurements of low velocity absorption during transit, 
and the time-varying geocoronal and interplanetary
emission in the line core that cannot be completely corrected for.

Absorption of Lyman-$\alpha$ photons during transit over the velocity range
--130 to +100 km~s$^{-1}$ raises the question of the origin of the 
high-velocity
hydrogen that is difficult to explain by a thermal plasma at moderate 
temperature. \citet{Holmstrom2008} proposed that the 
high-velocity neutral hydrogen can be produced  by charge exchange
between stellar wind protons and neutral hydrogen in the planet's outflow. 
Such energetic neutral atoms (ENAs) are seen in the solar system and 
likely occur where the stellar and planetary winds interact between 
HD~209458b and its host star. If this were the only possible explanation, then
the Lyman-$\alpha$ profile during transit
provides useful information on the stellar wind but ambiguous information on 
mass loss from the planet \citep{Holmstrom2008, Murray-Clay2009}. However,
\citet{Lecavelier2008b} argued that the observed Lyman-$\alpha$ profile 
and the high-velocity hydrogen atoms can be simply explained by stellar 
radiation
pressure and the ENA explanation required a peculiar stellar wind model.

The Lyman-$\alpha$ transit observations by 
\citet{Vidal-Madjar2003, Vidal-Madjar2004} stimulated a number of interpretive
papers concerning the mechanisms for mass loss and lifetime of a Jupiter-like 
planet close to its host star [e.g., \citet{Vidal-Madjar2003, Baraffe2004,
Koskinen2007, Holmstrom2008, Koskinen2010}] and hydrodynamic models 
[e.g., \citet{Lammer2003, Lecavelier2004, Yelle2004, Jaritz2005, Tian2005, 
Garcia2007, Schneiter2007, Murray-Clay2009}].
As discussed below, the theoretical models differ in their treatment of 
the amount of available radiative energy input 
(photoionization and otherwise) to 
drive the hyrdodynamic blow-off, the location of the heating, the presence
of metals in the outflow, and the relative 
amount of cooling by expansion, thermal conduction, and radiation. These
models predict very different mass-loss rates. Which
of these models are consistent with the spectrally resolved metal lines 
observed during transit?

With the objectives of measuring an accurate mass-loss rate from the 
planet and testing different physical models for the mass loss, one needs
far better data, which requires a new scientific instrument. 
One needs higher S/N spectra with higher spectral resolution 
and sufficient sensitivity to study many 
spectral lines formed in the extended atmosphere and wind of the exoplanet.
In particular, it is important to use spectral lines other than Lyman-$\alpha$
to avoid unique difficulties in the analysis of this line.
As we describe below, the new Cosmic Origins Spectrograph (COS) on {\it HST}  
is well designed for this task.

\section{OBSERVATIONS AND DATA REDUCTION}

Installed on the {\it HST} during Servicing Mission 4 in 2009 May, COS is a 
high-throughput ultraviolet (UV) spectrograph optimized for point sources. 
Descriptions of the on-orbit performance characteristics of COS
will be presented by Green et al. (2010, in preparation) and
Osterman et al. (2010, in preparation). 
During GTO Program 11534, we observed HD~209458 with both the G130M and G160M 
gratings of the COS far-UV channel to obtain moderate-resolution
spectra covering the 1140--1790~\AA\ spectral region.
In this paper, we analyze the spectra obtained with the G130M grating, which
includes the spectral range 1140--1450~\AA.
The G160M observations are presented by \citet{France2010}. 
The COS line-spread function (LSF) in the far-UV is not a simple Gaussian, 
but it 
can be approximated by a Gaussian with a resolution of 17,000-18,000 
(16.7-17.6 km~s$^{-1}$) with extended wings \citep{Ghavamian2010}.

We have tested for the effects of the extended wings by comparing the
C~II 1335.6854~\AA\ emission line of $\alpha$ Cen~A (G2~V) 
observed with the STIS E140H
grating (3 km~s$^{-1}$ resolution) convolved with a Gaussian (17 km~s$^{-1}$
resolution) and with the COS LSF. In the line wings at 50--70 km~s$^{-1}$, the
Gaussian broadens the profile by less than 1 km~s$^{-1}$, 
but the COS LSF broadens the
profile by 2--6 km~s$^{-1}$. Since both the transit and non-transit line 
profiles are broadened in the same way, 
their ratio is not changed by the broad wings of the COS LSF. The broadening
of the intrinsic spectral lines by the COS LSF (compared to the Gaussian 
profile) is less than 1/3 of the spectral resolution element and thus not 
significant.

Table 2 summarizes the individual G130M exposures obtained at transit 
(near orbital phase 0.00), first quadrature (phase 0.25), secondary eclipse
(phase 0.50), and second quadrature (phase 0.75). 
The total exposure times for these phases are 4946.8 s, 7941.7 s, 4956.7 s,
and 7796.0 s, respectively. The orbital phases 
listed in Table~2 are computed using the ephemeris of \citet{Knutson2007}. 
Transit occurs between 
orbital phases 0.982 and 0.018 \citep{Wittenmyer2005} as indicated by the
optical light curve. Since the duration of transit is about 3.22 hours,
{\it HST} could observe HD 209458 during the same transit 
on two sequential spacecraft orbits. 

The S/N in the COS spectra can be lower than predicted by photon 
statistics as a result of two instrumental effects. The hex pattern 
formed at the intersection of the microchannel plate pores 
imposes a $\pm 4$\% 
pattern on the spectral signal, and shadows of the wire grid 
on the detector decrease 
the signal as much as 20\% at certain locations on the detector faceplate.
We checked the locations of the known wire grid shadows and found that 
the spectra of the C~II, Si~III, and Si~IV lines studied in this paper 
are not affected by the wire grids.
Both effects, which are fixed at the detector, were mitigated by 
moving the grating and the resultant spectrum to different positions 
on the detector. In this program, we observed at four different 
grating offset positions identified by the beginning
and ending wavelengths of the gap between the two detector 
faceplate elements listed in
the last column of Table~2. In principle, division of a transit spectrum by 
an non-transit spectrum at the same grating position should cancel the
signal errors introduced by both instrumental 
effects to show the true spectral difference produced
by the planet in front of the star. In practice, the cancellation 
will not be perfect because of possible 
time variations in the stellar spectrum
and grating mechanisms that do not return to the same 
position every time. The grating mechanism positioning errors 
are not fully understood at this time. We found that the wavelength 
solutions for 
spectra observed at different grating offset positions are not identical 
and could be off by 1--2 pixels. 
We registered each observed spectrum with others 
at the same grating setting by cross-correlating the interstellar feature 
in the C~II 1334.5323~\AA\ line. These registrations typically 
involved displacements by only 1--2 pixels, far less than a resolution
element with no significant degradation in the spectral resolution.
We then co-added the summed spectra for the four grating offset positions 
using the same cross-correlation technique.
Also, within 500 pixels of the end of each detector
segment (about 5~\AA\ for G130M), 
there are additional instrumental artifacts in the spectrum. For
this reason, we did not analyze portions of the spectrum 
near the ends of a detector faceplate. 
The data were processed with the COS calibration pipeline,
CALCOS\footnote{We refer the reader to the cycle 18 COS Instrument Handbook
for more details: http://www.stsci.edu/hst/cos/documents/handbooks/current/cos\_cover.html.} 
v2.11b (2009-09-08) and combined with a custom IDL coaddition procedure. 

Figure 1 shows a portion of the G130M spectra obtained during
transit and all non-transit phases
(secondary eclipse and both quadratures). The spectra are very 
similar, but there are subtle differences that provide information
on the exoplanet's atmosphere and mass-loss rate. 
In this paper, we compare transit and 
non-transit spectra of lines of C~II, Si~III, and Si~IV. The H~I Lyman-$\alpha$
line cannot be analyzed because geocoronal emission through the large 
aperture of COS completely dominates any stellar componant. The O~I lines at 
1302, 1304, and 1306~\AA\ were also not analyzed because time-varying 
geocoronal emission in these lines renders comparison of transit and
non-transit spectra uncertain and because the lines were located close 
to the end of a detector faceplate.

\section{RESULTS}

\subsection {\it Si~IV Line}

A critical question is whether the emission line fluxes of the host star
were significantly different at the times when the transit and non-transit 
spectra were obtained, since late-type stars like the Sun often show 
time-variable emission in lines formed in their outer atmospheres 
\citep[e.g.,][]{Rottman2006}. 
Figure 2 shows the co-added transit and non-transit spectra for the
Si~IV resonance line (1393.76~\AA). 
Since the Si~IV line is formed in highly ionized gas which is
not likely to be in the atmosphere or extended wind of the exoplanet,
the very similar Si~IV fluxes during times of transit and non-transit 
indictate that 
the average flux in the stellar emission lines during our observations
was nearly constant. The difference spectrum panel in Figure 2 confirms that
the transit and non-transit spectra of the Si~IV line are identical to within
the noise level. 

To make the comparison of the transit and non-transit Si~IV line profiles more
quantitative, we compute the mean transit/non-transit
flux ratio over the velocity range --50 to +50 km~s$^{-1}$. Beyond this 
velocity range, the fluxes are less than 15\% of the peak flux, and the 
S/N in each spectral resolution element is small. For the 
100 km~s$^{-1}$ wide velocity interval centered on 0 km~s$^{-1}$, the mean flux
ratio is 0.998. To estimate the error in the mean flux ratio, we have computed
flux ratios for many velocity intervals contained
within the --50 to +50 km~s$^{-1}$ velocity range that are between 
60 and 95 km~s$^{-1}$ wide. 
The dispersion of these flux ratios indicates that the standard deviation 
in the mean flux ratio for the --50 to +50 km~s$^{-1}$ velocity range 
is $\pm 0.014$.
We therefore conclude that the mean stellar fluxes in the other emission lines 
were the same during transit and non-transit times, 
and do not rescale the transit and non-transit fluxes. At positive velocities,
there are regions where the transit/non-transit ratio may differ from unity at 
or above one standard deviation. We discuss this in Section 3.4.

\subsection{\it C II Lines}

The C II lines at 1334.5323~\AA\ and 1335.6854~\AA\ are bright emission lines
formed in the chromospheres of solar-type stars like HD~209458.  
On October 2, 2009, we observed the C~II lines during transit with the G130M 
grating at four grating positions specified by the gap wavelengths listed 
in Table~2. For each of these grating positions, there are corresponding 
observations at both quadratures and a secondary eclipse 
obtained at different times. 

Figure 3 shows the co-added spectra of the C~II 1334.5323~\AA\ resonance line
and the C~II 1335.6854~\AA\ line. The spectra were smoothed with a 
5-pixel boxcar, which is somewhat smaller than the on-orbit G130M spectral 
resolution element of 7--8 pixels.
The transit and non-transit spectra are plotted 
separately and ratioed. The deep minimum observed in the C~II 1334.5323~\AA\ 
line at --6.60 km~s$^{-1}$ (heliocentric) is interstellar absorption due 
to the Eri cloud located within 3.5 pc of the Sun \citep{Redfield2008}.
Since the radial velocity of HD~209458
is --14.8 \kms, the --6.6 km~s$^{-1}$ interstellar absorption is centered 
to the red of the centroid of the stellar emission line.

Both C~II lines show less flux during transit than non-transit times, 
although there are small velocity intervals where the transit/non-transit
ratio exceeds unity. To test whether the ratios in these velocity 
intervals are real or noise, 
we have co-added the profiles of the two C~II lines (including the portion 
of the 1334~\AA\ line with interstellar absorption) to get transit
and non-transit profiles with twice the signal and higher S/N. The co-added 
C~II profiles are included in Figure 3, together with difference and 
ratio profiles for each C~II line and the co-added line. The co-added C~II
line shows absorption during transit at all velocities between --50 and +50
km~s$^{-1}$, except at zero velocity. The deep absorption features for the
co-added C~II line are interesting and will be discussed in Section 3.4.

Using the same method that we used for the Si~IV line, we find that the 
mean flux ratio in the --50 to +50 km~s$^{-1}$ velocity interval is
$0.924\pm 0.022$ for the C~II 1334.53~\AA\ line and $0.921\pm 0.015$ for the
C~II 1335.69~\AA\ line. For the co-added C~II line, the mean flux ratio
is $0.922\pm 0.013$.

\subsection{\it Si III Line}

Figure 2 shows the co-added transit and non-transit spectra for the 
Si~III resonance line at 1206.500~\AA. The difference and ratio spectra
show absorption at all velocities between -60 and +60 km~s$^{-1}$.
There are also deep absorption 
features which we will discuss in Section 3.4.

Using the same method that we used in
analyzing the Si~IV and C~II lines, 
we find that the mean flux ratio in the --50 to +50 km~s$^{-1}$ band is
$0.918\pm 0.014$, the same within errors as previously seen for the C~II
lines. Our result for the mean absorption 
depth during transit of $8.2\pm 1.4$\%  is very different 
from the non-detection absorption depth of $0.0^{+2.2}_{-0.0}$\% obtained by 
\citet{Vidal-Madjar2004} from the STIS G140L spectrum and confirmed by
\citet{Ben-Jaffel2010}. 
We suspect that the difference between our detection of Si~III absorption 
during transit and the previous non-detection is due to time variability.
Charge transfer between protons and Si$^+$ ions leads to a substantially higher
abundance of Si$^{++}$ than is the case for collisional ionization equilibrium.
\citet{Baliunas1980} found that Si$^{++}$ is the dominant ionization stage in
coronal plasmas at temperatures above 20,000~K and is important at 15,000~K.
Thus, we are not surprized to see a substantial amount of 
Si$^{++}$ at the limb of the exoplanet's outflow where protons 
from the solar wind and planetary escape
are present to ionize Si$^+$, but the amount of Si$^{++}$ could vary 
appreciably with changes in the stellar wind, planetary mass-loss rate,
and temperature of the gas escaping from the planet's exosphere.

\subsection{\it Is There Velocity Structure in the Planet's Mass Loss?}

Until now we have discussed the mean transit/non-transit flux ratios for the
entire line cores between --50 and +50 km~s$^{-1}$, but there are indications 
of departures from these mean ratios which indicate that the absorption 
in the planet's mass loss depends on velocity. To test this idea, we 
compare in Figure~4 the difference and ratio profiles for the 
Si~III and co-added C~II lines. We find deep absorption features
near --10 and +15 km~s$^{-1}$ in both lines that are larger than the 
indicated errors shown in Figures~2 and 3. 
Since the C~II and Si~III lines are likely formed in the same region of the
exosphere, we include in Figure~4 the co-addition of the two profiles with 
typical error bars for comparison with theoretical models.
The increase in S/N with the co-addition of the C~II and Si III spectra allows
us to make more definitive statements about the velocity structure.
The dips near --10 and +15 km~s$^{-1}$ are more than twice as large as the 
errors per spectral resolution element (about 17 km~s$^{-1}$) 
and are most likely real. In addition, the ratios near --40 and +30--70
km~s$^{-1}$ are also low, but the line fluxes decrease rapidly away from 
line center and these low ratios are less certain. 
Future observations with higher S/N are needed to 
confirm this first detection of velocity structure in the 
expanding atmosphere of an exoplanet. 
We also note that there are absorption 
features near +20 and +40 km~s$^{-1}$ in the transit/non-transit ratio for the
Si~IV line which are greater than the noise level. Future observations are 
also needed to test the reality of these features. 

\subsection{\it Does HD 209458b have a Comet-like Tail?}

\citet{Schneider1998} proposed that giant exoplanets located close 
to their host
star will have comet-like tails of ionized gas produced by the interaction
of the stellar wind and ions excaping from the planet's exosphere, and that
these tails could produce absorption when observed in front of the star.
\citet{Vidal-Madjar2003} then predicted that stellar Lyman-$\alpha$ 
radiation pressure on hydrogen atoms escaping from the planet's exosphere 
would form a comet-like tail trailing the planet. 
The presence of gas in an extended tail should produce more absorption 
at and beyond egress than during ingress phases. 
\citet{Schneiter2007} calculated 
possible cometary wakes and the time evolution of Lyman-$\alpha$ absorption
from ingress to beyond egress. We searched our C~II and Si~III spectra 
for line profile differences
between the ingress and egress phases to see whether these ions could be 
present in significant amounts in a tail, but we found no differences larger 
than the noise level. 
Future observations with higher S/N are required to verify the predictions
of comet-like tails.

\section{MASS-LOSS RATE IN A THERMAL PLASMA OUTFLOW}

The $\approx8\pm 1$\% obscuration of the star in the C~II and Si~III lines
during transit far exceeds the 1.5\%
obscuration by the exoplanet's disk, indicating absorption by an extended
atmosphere or mass loss that is optically thick in the C~II and Si~III lines.
Our result for the C~II lines is consistent with the  
$7.5^{+3.6}_{-3.4}$\%  absorption depth that \citet{Vidal-Madjar2004} 
obtained for the unresolved C~II lines, but very different from 
the absorption depth of $0.0^{+2.2}_{-0.0}$\% 
that they obtained for the Si~III line. Their absorption depth 
of $0.0^{+6.5}_{-0.0}$\% for the Si~IV line is consistent with our result.

Since the 
size of the obscuring material (optically thick in these lines) is
similar to the size of the exoplanet's Roche lobe \citep{Ben-Jaffel2010},
 one possible explanation for
the transit data is that the Roche lobe is filled by mass loss from the planet.
Another possible explanation is that the exoplanet's escaping gas
forms an extended cometary tail blown out by the stellar wind, 
as suggested by hydrodynamical simulations \citep{Schneiter2007}. 

We estimate the mass-loss rate from the planet's atmosphere by assuming that 
the outflow forms a spherically symmetric envelope around the planet which 
is optically thick in the C~II resonance line covering about 8\% 
of the stellar surface. The spherical symmetry assumption is a good 
approximation because the shape of the tidally induced Roche lobe as seen 
during transit is close to a sphere \citep{Lecavelier2004}.
The envelope becomes optically thin at its ``limb''
because gas expansion and mass flux conservation  
reduce the gas density with increasing radial distance ($r$)
from the planet.
We define the envelope ``limb'' to be at the radial distance from the planet 
for which the absorption
area is the same as that of an opaque sphere of the same size. 
 
We consider a line of sight from the star that 
passes through the envelope with $p$ the point of closest approach 
to the planet. The quantity $x$ measures the distance along this line of sight 
from point $p$ back to the star. 
The optical depth to the observer in the C~II resonance line along this line 
of sight at the apparent limb of the envelope is,

\begin{equation}
\tau(p,\Delta\nu) = 2k\int_{0}^{\infty} n_{C~II} (x) dx = 0.69,  
\end{equation}

\noindent where $n_{C~II}$ is the number density of C$^+$ ions,
$k = \frac{\sqrt{\pi} e^2}{mc}\frac{f}{\Delta\nu_{D}}$ 
is the line opacity,
$f=0.128$ is the oscillator strength, $m$ is the electron mass, and
$\Delta\nu_{D}$ is the Doppler width. 
The value of $p$ depends somewhat on the displacement from line center, 
$\Delta\nu$, because for expanding gas, most of the absorption 
occurs near the frequency corresponding to the line-of-sight 
component of the expansion velocity. The limb occurs at 
$\tau=0.69$,
where as much stellar flux is transmitted through the envelope for lines
of sight passing inside of $p$ as for those lines of sight 
passing outside of $p$, 
$\int_{0.69}^{\infty}e^{-t}dt = \int_0^{0.69}e^{-t}dt$, where $t$ is the
optical depth through the envelope for lines of sight with different values 
of $p$. Since the gas density and thus optical depth increase along lines of 
sight inside of $p$, the limb will occur at somewhat smaller optical depths 
depending on details the density and outflow velocity. For example, 
\citet{Lecavelier2008a} calculated $\tau=0.56$ for an
atmosphere in hydrostatic equilibrium, which has a density that decreases 
exponentially outward but does not have an outflow or heating in the outer 
layers. If the density decreases
with radial position according to a power law, then $\tau$ could be smaller 
than 0.69 by a factor of 2 or 3. The resulting mass loss rates would also 
be smaller by this factor. We call attention to this uncertainty that should
be with resolved by realistic models based on physical principles that 
predict transit line profiles consistent with the COS data. 

Since the gas density decreases with increasing radial distance from the 
planet $r$ (thus increasing $x$ along the line of sight), 
$p$ should decrease slightly with increasing $\Delta\nu$. 
We compute $\Delta\nu_D = 2.7 \times 10^{10}$ s$^{-1}$, 
assuming a gas temperature of 10,000~K \citep{Garcia2007, Ballester2007}. 
If the outflow
is turbulent or some absorbers have superthermal velocities (see below), 
then $\Delta\nu_D$ will be larger. 
For a spherical outflow with constant mass flux,  
the mass flux of C$^+$ ions from the planet is

\begin{equation}
\dot{M}_{C~II} = 4\pi r^2 m_C v n_{C~II}(r),
\end{equation}

\noindent where $v$ is the outflow speed at $r$,
$m_C$ is the mass of a carbon atom, and $r^2 = x^2 + p^2$.

We estimate $\dot{M}_{C~II}$ by requiring that the gas in the exosphere 
obscure 8\% of the stellar disk within $\Delta\nu_D$ of line center, 
corresponding to $\pm 3.7$ km~s$^{-1}$. This condition will
occur when the optical depth is about 0.69 
at $p^2=0.08 R^2_{\star}$, i.e.,

\begin{equation}
\tau(p=0.284R_{\star},\Delta\nu) = \frac{2\sqrt{\pi} e^2 f \dot{M}_{C~II}}
{mc\Delta\nu_D 4\pi m_C v}\int_0^{\infty} \frac{dx}{r^2} = 0.69.
\end{equation}

\noindent The integral has an analytical solution by setting $u=x/p$ of

\begin{equation}
\int_0^{\infty}\frac{dx}{r^2} = 
\frac{1}{p} \int_0^{\infty}\frac{du}{u^2 +1} = \frac{\pi}{2p}.
\end{equation}

\noindent We adopt $v=10$ km~s$^{-1}$ at $p=0.284R_{\star}=2.36R_{planet}$
from the theoretical models of \citet{Tian2005}, \citet{Garcia2007}, 
and \citet{Murray-Clay2009}. We then find that

\begin{equation}
\dot{M}_{C~II}=2.1 \times 10^7 \rm{g~s}^{-1} = 6.8 \times 10^{14} 
\rm{g~yr}^{-1}. 
\end{equation}

\noindent In his model for the exosphere of HD~209458b, \citet{Garcia2007} 
found that essentially all of the carbon is singly ionized. 
If we assume that the carbon abundance is solar, $2.7 \times 10^{-4}$ that 
of hydrogen \citep{Asplund2009}, the total mass-loss rate from the planet 
for only thermal broadening of the C~II lines is

\begin{equation}
\dot{M}_{total} = 8.0 \times 10^{10} \rm {g~s}^{-1} =
2.4 \times 10^{18} \rm{g~yr}^{-1}. 
\end{equation}

Since the planet's mass is $1.2 \times 10^{30}$ g \citep{Knutson2007}, 
the fractional mass-loss rate per year is very small, as shown by

\begin{equation}
\frac{\dot{M}_{total}}{M_{planet}} = 2.0\times  10^{-12} \rm{yr}^{-1}. 
\end{equation}

\noindent This calculation assumes that the mass-loss rate and 
planetary orbit do not change with time. Since additional line broadening 
is required to explain the broad absorption in the C~II lines, these values
of $\dot{M}_{C II}$ and $\dot{M}_{total}$ are at the lower end of their 
realistic ranges (see Section~5).

\section{DISCUSSION}

We now compare our empirical mass-loss rate with those of various 
theoretical models. All theoretical models of which we are aware 
assume hydrodynamical outflow driven by
high temperatures and pressures near the base of the outflow. The 
important questions are the amount and location of energy absorbed 
from the star, the important
cooling mechanisms, and the location of the sonic point in these transonic 
(Parker-like) winds relative to the location of the exoplanet's Roche lobe.
\citet{Vidal-Madjar2003} demonstrated that neutral hydrogen must be escaping 
from a volume larger than the planet's Roche lobe by an unspecified 
mass-loss mechanism 
and that stellar radiation pressure further accelerates the outflow   
and shapes a comet-like tail. They proposed that the mass-loss rate
is $\dot{M}> 10^{10}$ g~s$^{-1}$, but they added that 
``owing to saturation effects
in the [Lyman-$\alpha$] absorption line, a flux larger 
by several orders of magnitude would 
produce a similar absorption signature.'' \citet{Murray-Clay2009} computed
models with and without radiation pressure and concluded that radiation
pressure does not increase $\dot{M}$, but it can change the location of the
mass loss from the day side to the night side of the exoplanet. 
Tidal forces also do not change $\dot{M}$ appreciably, 
but they move the sonic point closer to the exoplanet 
and increase the flow velocity \citep{Garcia2007, Murray-Clay2009}.

All of the theoretical models assume that HD~209458 is a main sequence star
like the Sun with ultraviolet and extreme ultraviolet radiative output
similar to the Sun at low or high activity. 
Since mass-loss rates are proportional to the energy input 
(the so-called energy-limited escape rate), the critical question
is the energy input rate. In all of the models, expansion (PdV work) 
is the dominant cooling mechanism, 
although radiation by H$_3^+$ at the base of the flow must be included 
\citep{Yelle2004}, and radiation in the Lyman-$\alpha$ line
can be important for T Tauri star radiative input rates. 
Many models assume that only the energy remaining after stellar Lyman 
continuum (E$>13.6$ eV) photons ionize neutral hydrogen is available to 
energize the outflow, so that the 
assumed fraction of this radiation (83\% or 100\% in different models) 
controls the outflow. For such models, $\dot{M}$ is a few times $10^{10}$
g~s$^{-1}$. For example, \citet{Yelle2004}, as corrected in \citet{Yelle2006}, 
finds $\dot{M}=4.7\times 10^{10}$, 
\citet{Tian2005} find $\dot{M}<6\times 10^{10}$, \citet{Garcia2007} finds
$\dot{M}=6.1\times 10^{10}$ for EUV flux similar to the low activity Sun but
$\dot{M}=1.5\times 10^{11}$ for EUV flux like the active Sun.
\citet{Murray-Clay2009} find that $\dot{M}\sim 2\times 10^{10}$ 
g~s$^{-1}$.

The inclusion of atoms and molecules other than H, H$^+$, and H$_2$ 
can significantly change the outflow, in particular by increasing the 
mass-loss rate, because more of the stellar UV radiation
is available to heat the outflow and drive the mass loss. For example,
the Lyman-$\alpha$ line contains more flux than the entire Lyman continuum
\citep{Ribas2005}, and other UV emission lines and continuum radiation
in the wavelength range 912--2000~\AA\ can be absorbed in the bound-free 
continua of atomic species (e.g., CNO)
and molecules including H$_2$. When \citet{Garcia2007} added solar abundances 
of CNO to his models, $\dot{M}$ increased by a factor of 4.5 
(see his Table 6). Including partial absorption of the 1000--2000~\AA\ 
stellar flux by atoms and molecules with solar abundances increases 
$\dot{M}$ to $6.3\times 10^{11}$ g~s$^{-1}$. Metals must exist in
the distant outflow since we see absorption during transit 
in the C~II and Si~III lines. By equating the 
energy input of both Lyman continuum and Lyman-$\alpha$ with stellar fluxes
estimated from the rotation rate of HD~209458, \citet{Lecavelier2007}
found $\dot{M}\sim 3\times 10^{11}$ g~s$^{-1}$. Using a similar energy
balance argument, \citet{Lammer2003} obtained $\dot{M}\sim 3\times 10^{11}$
g~s$^{-1}$, but \citet{Yelle2004} suggested that this result could be an 
overestimate as a result of not including H$_3^+$ cooling and ionization of H.

A critical question is whether line absorption during transit is produced
entirely by the outflowing thermal plasma or whether high-velocity
atoms and ions (either at higher temperatures or nonthermal)
are required to explain the broadest part of the absorption line
profiles. \citet{Ben-Jaffel2010} show that neutral hydrogen absorption
by scaled hydrodynamic outflow models may explain the flux decrease 
during transit even at $\pm 100$ km~s$^{-1}$ from the center of the 
Lyman-$\alpha$ line, but such models predict much less absorption during 
transit than is detected in the unresolved O~I and C~II line flux measurements
\citep{Vidal-Madjar2004}. Additional high-velocity absorbers are needed
to explain the O~I, C~II, and probably also the Lyman-$\alpha$ data,
but there is no agreement concerning the source of the high-velocity
absorbers. One possibility is radiation pressure, but \citet{Murray-Clay2009}
argue that the observed absorption near +100 km~s$^{-1}$ from the center of
the Lyman-$\alpha$ line is difficult to 
explain by stellar radiation pressure as originally proposed by 
\citet{Vidal-Madjar2003}. However, radiation pressure on neutral
hydrogen atoms must be present, and \citet{Lecavelier2008b} argue that
radiation pressure by itself can explain the Lyman-$\alpha$ observations.
\citet{Holmstrom2008} and \citet{Murray-Clay2009} argue that the observed
decrease in Lyman-$\alpha$ flux near and beyond $\pm 100$ km~s$^{-1}$ 
during transit requires high-velocity H atoms that could be explained by 
charge exchange between stellar wind protons and neutral H atoms in
the exoplanet's outflow. These high-velocity H atoms are called ENAs
(energetic neutral atoms). 

\citet{Ben-Jaffel2010} have tested whether the presence of high-velocity 
atoms and ions could explain the flux reduction observed 
in the unresolved O~I and C~II line transit data. 
They assume that a portion of the gas has an effective temperature (either 
thermal or nonthermal) many times larger than the thermal temperature in
outflow models. For the unresolved C~II 1334, 1335~\AA\ multiplet, they can 
explain the observed flux reduction by C$^+$ ions with effective temperatures 
5.5--107 times larger than the background gas temperature ($T_B$) 
in typical hydrodynamic models 
depending on the properties of the outflow models and the assumed 
location of the C$^+$ ions. In Figure~9 of their paper, they show examples
of predicted C~II 1334~\AA\ line profiles during transit with 
$T_{CII}/T_B \sim 16$ and $\sim 28$ that can explain the total decrease 
in unresolved line flux during transit, but these particular examples
are either too narrow or triangular in shape, 
unlike the observed C~II line profiles which show 
significant flux reduction between --50 and +50 km~s$^{-1}$ with probable 
enhanced absorption near --10 and +15 km~s$^{-1}$. Ben-Jaffel
(private communication) informs us that other models with somewhat larger 
$T_{CII}/T_B$ ratios can explain the flat reduction in C~II line flux 
between --50 and +50 km~s$^{-1}$. However these models have {\it ad hoc} 
velocity distributions and they do not explain the observed velocity structure.
In their paper, they conclude that COS spectra
are needed to ``...reveal the true balance between thermal and non-thermal 
populations...''. We encourage the development of new models, 
preferably based on plausible physical mechanisms, to explain the 
observed line profiles.

Our analysis of the C~II line absorption during transit assuming only 
thermal absorption by 10,000~K gas and solar carbon abundance  
led to our estimate of
$\dot{M}_{total} = 8.0\times 10^{10}$ g~s$^{-1}$. This mass-loss rate
is somewhat larger than the theoretical estimates mentioned above that
assume an outflow with no metals. The assumption of no metals in the outflow 
is clearly unrealistic as
we observe absorption by C~II and Si~III at 2.36R$_{planet}$.
At the same time, our analysis of the C~II lines assuming only thermal line
broadening (at 10,000~K) is unrealistic as the thermal velocity 
is only 3.7 km~s$^{-1}$,
and we observe absorption out to $\pm 50$ km~s$^{-1}$. Since we measure the
mass flux at the limb of the outflow where the C~II optical depth is 
small ($\tau=0.69$), we cannot invoke optical thickness to explain the broad
profile. Note that the same is true for the Lyman-$\alpha$ line, but 
(as the referee has called to our attention) \citet{Ben-Jaffel2010} and  
other authors include opacity and Lorentzian line shapes at the limb as a 
possible mechanism for absorption line broadening.

Equation 3 shows that
$\dot{M}_{C II}$ is proportional to the Doppler broadening parameter
$\Delta\nu_D$. To obtain 8\% absorption at $\pm 50$ km~s$^{-1}$ from line
center requires either turbulent broadening with velocity of about 
50 km~s$^{-1}$ or superthermal C$^+$ ions with similar velocities. If 
turbulent broadening is responsible, then the mass-loss rate increases
by a factor of 13 to $\dot{M}_{total}=1\times 10^{12}$ g~s$^{-1}$. 
However, 50 km~s$^{-1}$ turbulent velocities are highly supersonic and
unlikely.
We estimate that a sensible upper limit to the turbulent velocity would be
twice the flow speed or 20 km~s$^{-1}$, for which 
$\dot{M}_{total}\approx 4\times 10^{11}$ g~s$^{-1}$. This would increase
$\dot{M}_{total}/M_{planet}$ to $\approx 4\times 10^{-11}$ yr$^{-1}$. 

If high-velocity C$^+$ ions are responsible for the broad absorption, 
then the mass-loss rate could be closer to the thermal value because only 
a small fraction of the C$^+$ ions are likely to have high velocities.
We conclude that $\dot{M}_{total}=$ (8--40)$\times 10^{10}$ g~s$^{-1}$ 
is consistent with the C~II line profiles. We find it 
interesting that when \citet{Garcia2007} includes metals with solar 
abundances in his hydrodynamic mass outflow models, 
the mass-loss rate increases from 
$\dot{M}=1.4\times 10^{11}$ to $\dot{M}=4.95\times 10^{11}$ g~s$^{-1}$, 
a value near the top end of our empirical mass-loss rate estimate.
Thus high-velocity C$^+$ ions are required to explain the 
C~II absorption during transit. Similar arguments would likely explain the
broad Si~III transit absorption.

\section{CONCLUSIONS}

The high sensitivity and moderate spectral resolution of COS allowed us to
obtain the first measurements of gas absorption in the exosphere 
of a transiting exoplanet in lines of C~II and Si~III with velocity 
resolution. During transits of HD~209458, we found that the mean flux in these
lines is reduced by $7.8\pm 1.3$\% for the co-added C~II lines and 
$8.2\pm 1.4$\% for the Si~III line 
in the velocity range --50 to +50~km~s$^{-1}$.
The 8\% absorption constraint sets the location of the outflow limb  
(2.36R$_{planet}$) where the C~II lines become optically thin. 
Theoretical hydrodynamic models show that the outflow velocity is about
10 km~s$^{-1}$ and gas temperature about 10,000~K at this location. 
If we assume that the
outflow is purely thermal, then $\dot{M}_{total}=8\times 10^{10}$
g~s$^{-1}$, somewhat larger than theoretical estimates based on outflows with
no metals. Since absorption in the C~II lines is much broader than
indicated by Doppler broadening with a thermal velocity of 3.7 
km~s$^{-1}$, the lines must be broadened by large turbulence or high-velocity
C$^+$ ions. If turbulent broadening dominates, then the 8\% absorption
seen out to $\pm 50$ km~s$^{-1}$ would require mass-loss rates as large as
$1\times 10^{12}$ g~s$^{-1}$, but such highly supersonic velocities are not 
likely. Instead, absorption by high-velocity C$^+$ and Si$^{++}$ ions formed 
by the interaction of the stellar wind and planetary mass loss or some other 
mechanism broadens the C~II
and Si~III lines. We conclude that the mass-loss rate
in the exoplanet's outflow is in the range 
(8--40)$\times 10^{10}$ g~s$^{-1}$. These rates assume that
the carbon abundance is solar, which is not the case for the giant planets
in the solar system. We note that 
mass-loss rates predicted by hydrodynamic models that include 
solar abundant metals in the outflow
lie at the upper limit of this range.

Comparison of the C~II and Si~III line depths and transit/non-transit line 
ratios shows deep absorption features near --10 and +15 km~s$^{-1}$ 
in both lines.
Since the C~II and Si~III lines are likely formed in the same region of the
exosphere, we include in Figure~4 the co-addition of the two profiles with 
typical error bars for comparison with theoretical models.
The increase in S/N with the co-addition of the C~II and Si III spectra allows
us to make more definitive statements about the velocity structure.
The dips near --10 and +15 km~s$^{-1}$ are more than twice as large as the 
errors per spectral resolution element (about 17 km~s$^{-1}$) 
and are most likely real. In addition, the ratios near --40 and +30--70
km~s$^{-1}$ are also low, but the line fluxes decrease rapidly away from 
line center and these low ratios are less certain. 
Future observations with higher S/N are needed to 
confirm this first detection of velocity structure in the 
expanding atmosphere of an exoplanet. 

The COS transit observations of the absorption velocity structure in 
these lines demonstrate the need for including high-velocity 
C$^+$ and Si$^{++}$ ions in models of the exosphere of HD~209458b and
presumably other close-in exoplanets. It remains to be seen whether the
inclusion of a simple distribution of ions with high effective temperature
as proposed by \citet{Ben-Jaffel2010} is adequate to explain the data. 
The new COS data provide the opportunity to test a range 
of  absorption models based on such physical mechanisms 
as radiation pressure, charge exchange, and
interactions between the stellar wind and the planet's outflow to understand
mass-loss processes in the exospheres of close-in planets.

This work is supported by NASA through grants NNX08AC146 and NAS5-98043
to the University of Colorado at Boulder. We thank the many people at the 
Space Telescope Science Institute, Goddard Space Flight Center, Ball 
Aerospace and Technologies Corp. and CASA at the University of Colorado for the
excellent hardware and software that has made COS a great success. We 
particularly wish to thank the referee for his many thoughtful and thorough 
comments and his bringing to our attention important publications that have 
led to significant modifications in the paper. We thank Dr. J. Schneider
for encouraging us to look for an assymetry between ingress and egress. We
also thank Dr. Lofti Ben-Jaffel for clarifying our understanding of the
Ben-Jaffel \& Hosseini (2010) paper. We thank Dr. Alain Lecavelier for calling
our attention to the different values of $\tau$ at the limb that depend on the
density structure in the envelope.

\begin{figure}
\includegraphics[angle=90, scale=0.70]
{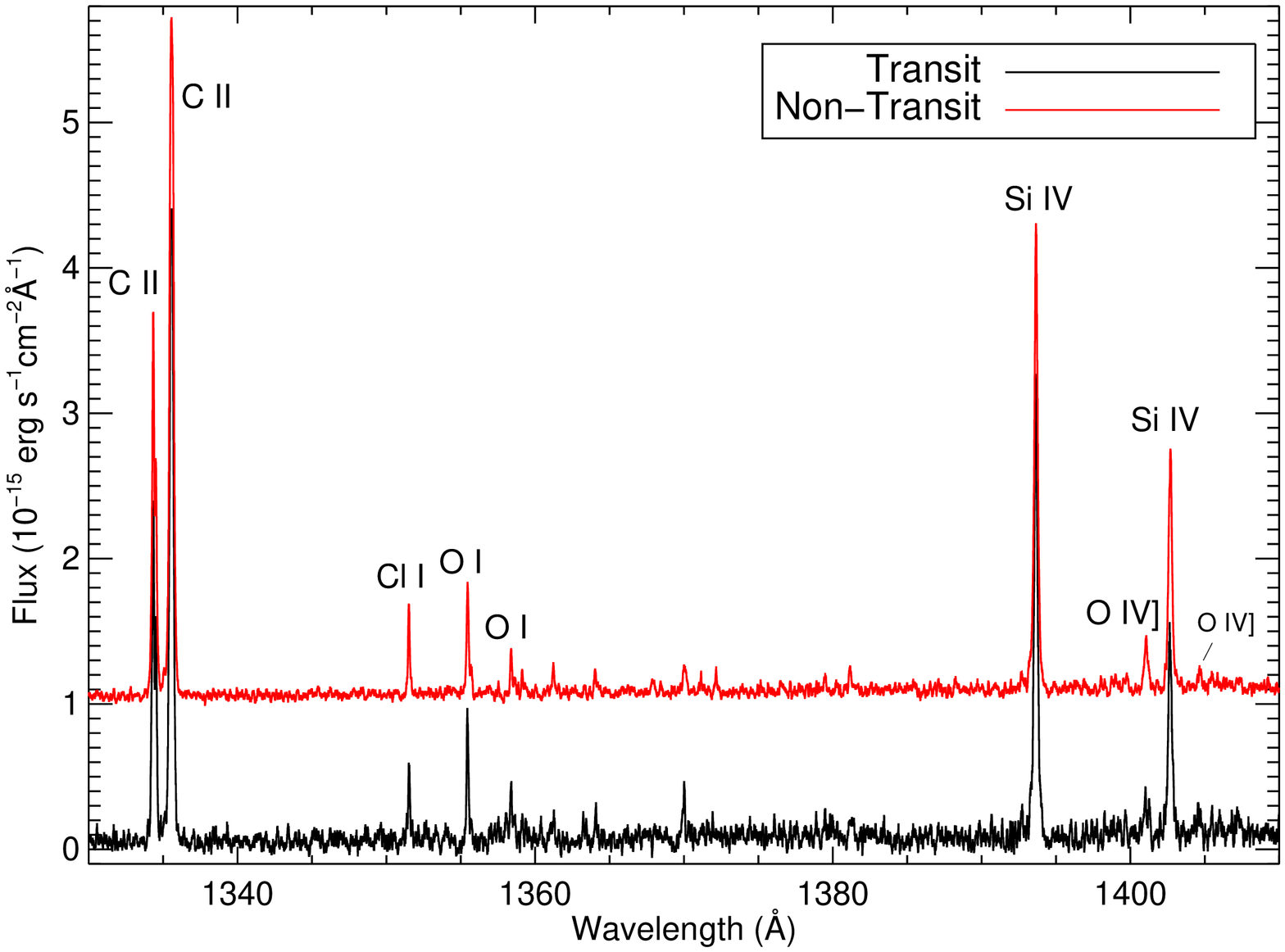}
\caption{A portion of the COS G130M spectrum of HD~209458. The lower (black)
plot is the sum of four spectra obtained during transit. The upper (red)
plot (displaced upward) is the sum of the ten quadrature and 
four secondary eclipse spectra.}
\end{figure}

\begin{figure}
\includegraphics[angle=90, scale=0.70]
{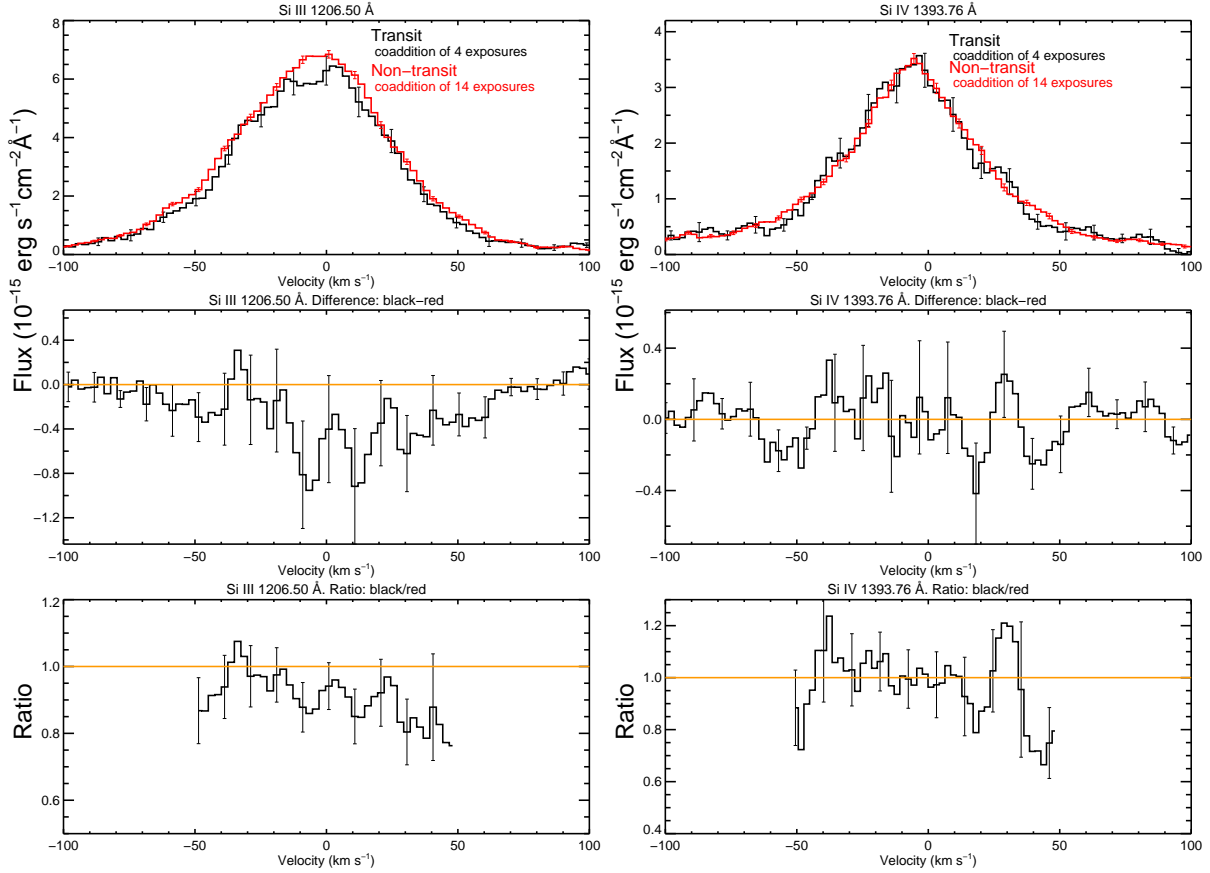}
\caption{Upper panels: Comparison of the four co-added spectra obtained 
during transit (black) with the co-added non-transit spectra 
(red) for the Si~III and Si~IV lines. 
Representative error bars per pixel are included. 
Middle panels: difference spectra (transit minus non-transit) with 
representative error bars.
Lower panels: flux ratio (transit/non-transit) spectra. The ratios are not 
plotted beyond $\pm 50$ km~s$^{-1}$ as the errors are large at low flux
levels. The velocity scale is relative to the --14.8 km~s$^{-1}$ 
radial velocity of the star.}
\end{figure}

\begin{figure}
\includegraphics[angle=90, scale=0.70]
{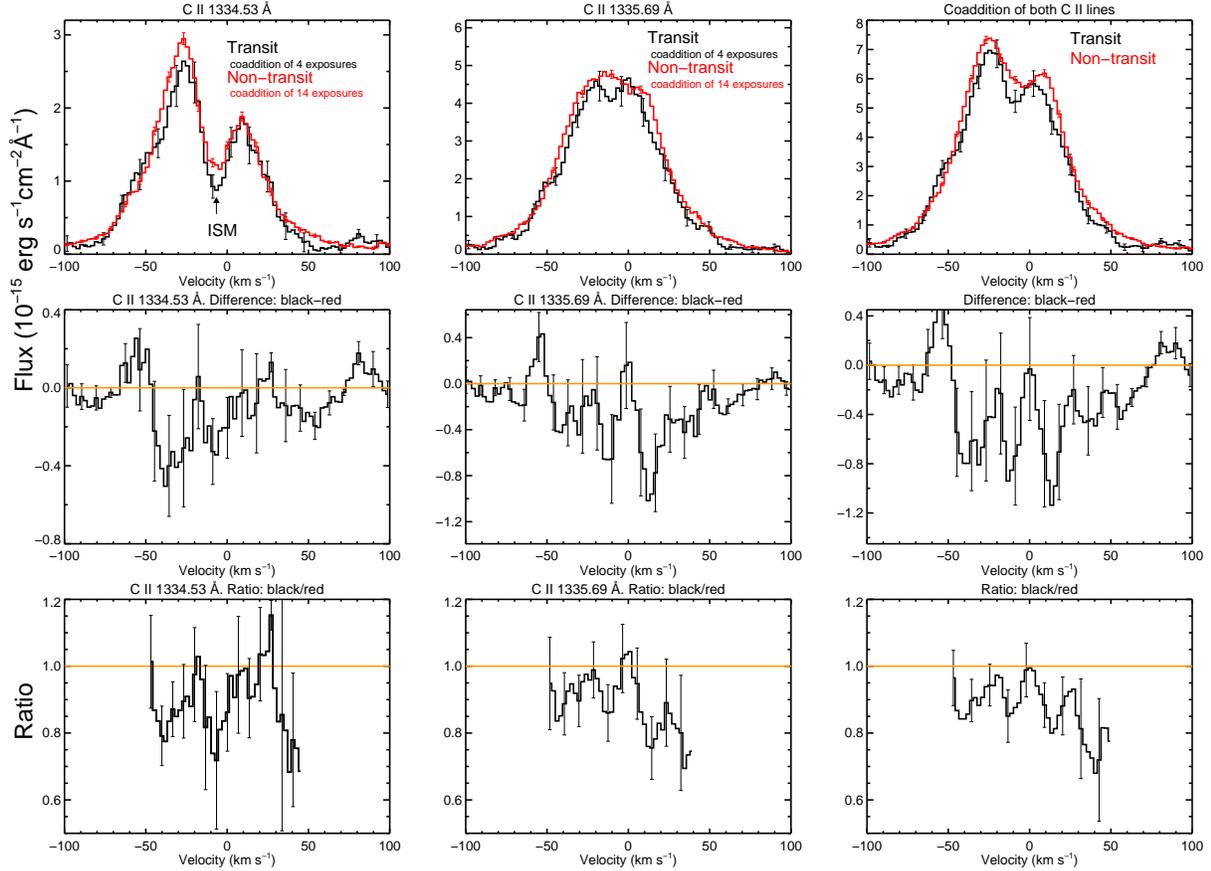}
\caption{Upper panels: Comparison of the four co-added spectra obtained 
during transit (black) with the co-added non-transit spectra 
(red) for the C~II 1334.53~\AA\ and 1335.69~\AA\ lines. The upper right
panel show a co-addition of both C~II lines on the same velocity scale. 
Representative error bars per pixel are included. 
Middle panels: difference spectra (transit minus non-transit) with 
representative error bars for each C~II line and the combined C~II line.
Lower panels: flux ratio (transit/non-transit) spectra. The ratios are not 
plotted beyond $\pm 50$ km~s$^{-1}$ as the errors are large at low flux
levels. The velocity scale is relative to the --14.8 km~s$^{-1}$ 
radial velocity of the star.}
\end{figure}

\begin{figure}
\includegraphics[angle=90, scale=0.70]
{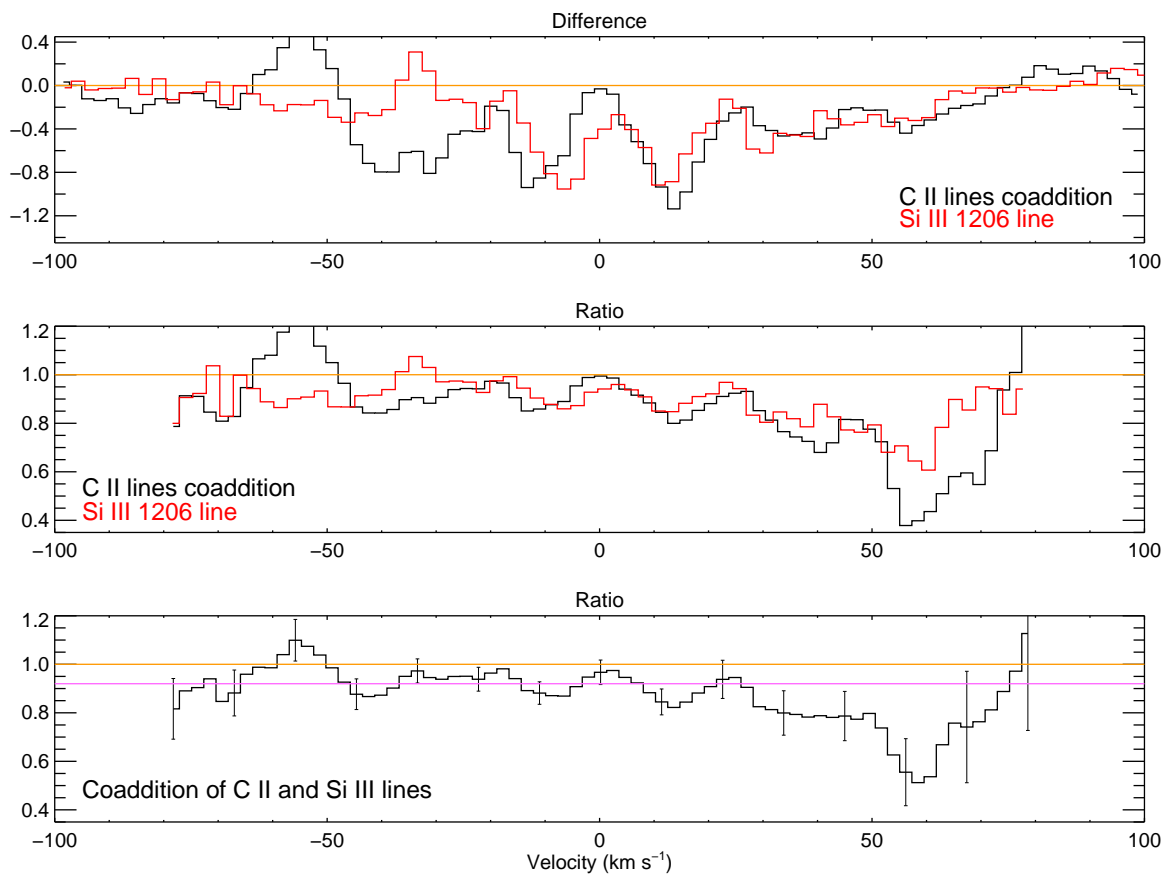}
\caption{Upper and middle panels: Comparison of the difference spectra 
(transit minus non-transit) and
ratio spectra (transit/non-transit) for the co-addition of the C~II Lines 
(black) and the Si~III line (red). 
Lower panel: Co-addition of the C~II and Si~III ratio spectra with typical
errors per spectral resolution element (about 17 km~s$^{-1}$). The horizontal 
line at 0.92 is the mean ratio for the velocity interval --50 to 
+50 km~s$^{-1}$. 
The dips near --10 and +15 km~s$^{-1}$ are each more than 2$\sigma$ and likely 
real. The low ratio features near --40 and +30--70 km~s$^{-1}$ occur where 
the line fluxes are low and are thus less certain.}
\end{figure}

\begin{deluxetable}{cc}
\tablewidth{0pt}
\tablenum{1}
\tablecaption{Properties\tablenotemark{a} of HD 209458 and HD 209458b}
\label{tab:prop}
\tablehead{Spectral type         & G0 V} 
\startdata
Distance (pc)                   & 47\\
$M_{\star}/M_{\odot}$           & $1.101^{+0.066}_{-0.062}$\\
$R_{\star}/R_{\odot}$           & $1.125^{+0.020}_{-0.023}$\\
$P_{\rm orbit}$ (days)          & 3.52474859(38)\\
$R_{\rm planet}/R_{\rm Jup}$    & $1.320^{+0.024}_{-0.025}$\\
$M_{\rm planet}/M_{\rm Jup}$    & $0.64\pm 0.06$\\
$\rho_{\rm planet}/\rho_{\rm Jup}$ & $0.26\pm 0.04$\\
Semimajor axis (AU)             & 0.045\\
Stellar flux at planet/Jupiter  & 13,400\\
Escape speed (km/s)             & $42\pm 4$\\
Orbital Velocity ampl. (km/s)   & 146\\
Transit duration (hr)           & 3.22\\
Transit depth for $R_{\rm planet}$ & 1.5\%\\ 
Transit depth for $R_{\rm Roche}$ & 10\%\\ \hline
\enddata
\tablenotetext{a}{Data from \citet{Knutson2007}.}
\end{deluxetable}

\begin{deluxetable}{cccccc}
\tablewidth{0pt}
\tabletypesize{\footnotesize}
\tablenum{2}
\tablecaption{COS G130M Observations of HD 209458 (Program 11534)
\label{tab:obs}}
\tablehead{Exp. Time (s) & Phase & Day & Start & Stop & 
Gap\tablenotemark{a} (\AA)}
\startdata


2340.2 &0.2516--0.2593&19 Sep & 10:10:15 & 10:49:15 & 1278--1290\\
0955.2 &0.2684--0.2716&19 Sep & 11:35:37 & 11:51:32 & 1288--1299\\
1851.2 &0.2722--0.2782&19 Sep & 11:54:40 & 12:25:31 & 1296--1306\\
0560.0 &0.2873--0.2891&19 Sep & 13:11:29 & 13:20:49 & 1288--1299\\
2235.1 &0.2898--0.2971&19 Sep & 13:24:06 & 14:01:21 & 1306--1316\\

2340.2 &0.7059--0.7136&24 Sep & 13:12:09 & 13:51:09 & 1278--1290\\
0945.2 &0.7229--0.7260&24 Sep & 14:38:04 & 14:53:49 & 1288--1299\\
1787.2 &0.7266--0.7324&24 Sep & 14:56:57 & 15:26:44 & 1296--1306\\
0925.2 &0.7417--0.7448&24 Sep & 16:13:54 & 16:29:19 & 1288--1299\\
1798.2 &0.7454--0.7513&24 Sep & 16:32:36 & 17:02:34 & 1306--1316\\


1045.2 &0.9913--0.9947&02 Oct & 14:32:10 & 14:49:35 & 1278--1290\\
1096.2 &0.9954--0.9990&02 Oct & 14:52:52 & 15:11:08 & 1288--1299\\
1400.2 &0.0077--0.0123&02 Oct & 15:55:28 & 16:18:48 & 1296--1306\\
1405.2 &0.0129--0.0175&02 Oct & 16:21:56 & 16:45:21 & 1306--1316\\


1057.2 &0.4880--0.4915&18 Oct & 10:57:04 & 11:14:41 & 1278--1290\\
1093.1 &0.4921--0.4957&18 Oct & 11:17:49 & 11:36.02 & 1288--1299\\
1401.2 &0.5043--0.5089&18 Oct & 12:19:37 & 12:42:58 & 1296--1306\\
1405.2 &0.5095--0.5141&18 Oct & 12:46:06 & 13:09:31 & 1306--1316\\

\enddata
\tablenotetext{a}{The gap is the wavelength interval between the two 
detector faceplate elements. No data are available for the gap region, and 
there are instrumental artifacts at wavelengths close to the gap.} 
\end{deluxetable}

\end{document}